\documentclass[12pt]{article}
\usepackage{moriond,psfig,epsfig,rotating}
\begin{document}

\newcommand{\lsim}   {\mathrel{\mathop{\kern 0pt \rlap
  {\raise.2ex\hbox{$<$}}}
  \lower.9ex\hbox{\kern-.190em $\sim$}}}
\newcommand{\gsim}   {\mathrel{\mathop{\kern 0pt \rlap
  {\raise.2ex\hbox{$>$}}}
  \lower.9ex\hbox{\kern-.190em $\sim$}}}
\def\be{\begin{equation}}
\def\ee{\end{equation}}
\def\ba{\begin{eqnarray}}
\def\ea{\end{eqnarray}}
\def\ap{\approx}
\title{Ultrahigh energy cosmic rays and supersymmetry%
\footnote{%
Talk given at {\em XXXIIIrd Rencontres de Moriond: Electroweak
Interactions and Unified Theories}}
}

\author{M. Kachelrie{\ss}}

\address{INFN, Laboratori Nazionali del Gran Sasso,
         I--67010 Assergi (AQ), Italy}

\maketitle

\begin{abstract}
Recently, models proposing superheavy particles $X$ as source of ultrahigh 
energy cosmic rays have attracted some interest. The $X$-particles are
either metastable relic particles from the early
Universe or are released by topological defects. In these models, 
the detected air-showers are produced by primaries originating from the 
fragmentation of the $X$-particles. We present the fragmentation
spectrum of superheavy particles 
calculated in SUSY-QCD. Then we discuss the status of the
lightest supersymmetric particle as possible ultrahigh energy primary.
\end{abstract}

\section{Introduction}

Cosmic rays \cite{GF} (CR) are observed in a wide energy range,
starting from subGeV  
energies up to $3\cdot 10^{20}$~eV. Apart from the highest energies, these 
particles are accelerated in our Galaxy, most probably by shocks produced 
by SNII explosions. There is no universal definition of Ultra High Energy 
Cosmic Rays (UHECR). We will use this term for energies
$E \gsim 10^{19}$~eV, 
where a new, flatter component appears in the CR spectrum (Fig.~1). The 
highest energies detected so far are $2-3\cdot 10^{20}$~eV.

It is natural to think that the UHE component has an extragalactic origin, 
since the galactic magnetic field cannot confine particles of these energies.
Moreover, the acceleration of protons or nuclei up to $2-3\cdot 10^{20}$~eV
is difficult to explain with the known astrophysical galactic sources. 
The most prominent signature of extragalactic UHECR is the so called
Greisen-Zatsepin-Kuzmin (GZK) cutoff~\cite{GZK}: 
The energy losses of protons, 
nuclei and photons sharply increase at $E_{\rm GZK} \sim 3\cdot 10^{19}$~eV,
reducing the mean free path length $l$ of these primaries to
less than 
$50$~Mpc or so. Thereby, the spectrum should become stepper above 
$E_{\rm GZK}$ for {\em any\/} source with a distribution homogeneous
on scales larger than $l$. 
There is another argument which disfavours the standard
astrophysical sources: At energies $E\sim 10^{20}$~eV, the arrival
direction of the primaries (which is known within several
degrees) should point towards their site of origin. But 
no source of UHECR like, e.g.,  active galactic nuclei has been
found within $50$~Mpc in the direction of these events.

An elegant solution to the above problems are top-down models: In
contrast to the standard sources, the primaries are not accelerated but
are the fragmentation products of some decaying superheavy particle
$X$. For $X$-particles with mass $m_X\gg E_{\rm GZK}$,
the acceleration problem is solved trivially. Moreover, 
these sources also evade detection by normal astronomical methods.

This contribution is organized as follows:
In Section 2, the in our view two most promising top-down models are
presented. In Section 3, the spectrum of hadrons produced in the
decay of supermassive particles is calculated in supersymmetric QCD.
As an application, the CR fluxes produced by cosmic necklaces are
shown in Section 4. Finally, the status of the LSP as UHE primary is
discussed in Section 5.

\begin{figure}
\psfig{bbllx= 110pt, bblly=350pt, bburx=500pt, bbury=600pt,
file=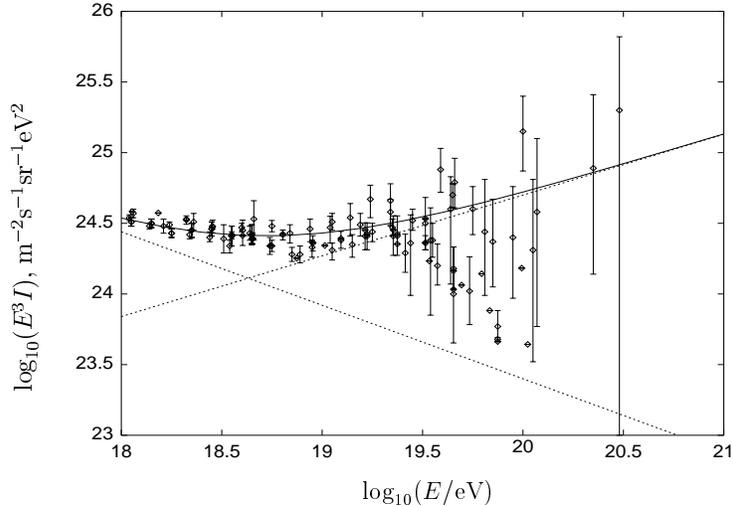, height= 7cm , clip=}
\caption{Compilation of the CR spectrum from AGASA, Fly's Eye, Haverah
Park and Yakutsk. A two component fit to the spectrum is also shown.}
\end{figure}

\section{Top-down models as new sources for UHECR}

Three different possibilities for top-down models have been discussed
in the literature: 

\begin{enumerate}
\item
Primordial black holes: During the final stage of their evaporation,
they emit high energy particles. However, when the resulting spectra
of cosmic rays  are combined with various observational 
bounds on the mass fraction of the universe in black holes, 
one finds that the UHE CR flux from black holes is well below the observed 
flux.

\item
Topological defects~\cite{td} such as superconducting
strings,  monopoles, and monopoles connected by strings: 
Here we will concentrate on {\em cosmic necklaces\/}, since this model 
seems to provide the largest UHE particle flux for fixed density of 
electromagnetic cascade radiation. 

Cosmic necklaces are hybrid defects
consisting of monopoles connected by a string. These defects are produced 
by the symmetry breaking $G\to H\times U(1) \to H\times Z_2$.
In the first phase transition at scale $\eta_m$, monopoles are 
produced. At the second phase transition, at scale  $\eta_s<\eta_m$, each 
monopole gets attached to two strings. The basic parameter for the evolution
of necklaces is the ratio $r=m/(\mu d)$ of the monopole mass $m$ and
the mass of the string 
between two monopoles, $\mu d$, where $\mu \sim \eta_s^2$ is the mass 
density of the string and $d$ the distance between two monopoles.
Strings loose their energy and can contract due to gravitational radiation.
As a result, all monopoles annihilate in the end producing
superheavy Higgs, gauge bosons and their supersymmetric partners 
which we call collectively $X$-particles.
The rate of $X$-particle production can be estimated as
\be
 \frac{d n_X}{d t} \sim \frac{r^2 \mu}{t^3 m_X} \,.
\label{n/t}
\ee
The flux of UHECR is determined mainly by two 
parameters, $r^2\mu$ and $m_X$, which values must be of order 
$10^{27}$~GeV$^2$ and $10^{14}$~GeV, respectively, to have 
the flux close to the observed one.
For a more complete discussion see Ref.~[4].   

\item
Superheavy, metastable relic particles $X$: They constitute
(part of) the cold dark matter (CDM) and, consequently, their abundance in
the galactic halo is enhanced by a factor $\sim 5\times 10^4$ above
their extragalactic abundance. 
Therefore, the proton and photon flux is dominated by the halo
component and the GZK-cutoff is avoided as was first pointed out
in Ref.~[5]. 

The necessary lifetimes, $\tau\gsim 10^{17}$s, and mass ranges, $m_X \gsim
10^{13}$~GeV, of the $X$-particle arise quite naturally in several
extensions of the
standard model. The $X$-particle could be protected by some sort of
$R$-parity which is extremely weakly broken by wormhole \cite{bkv97} or
instanton \cite{kr97} effects, or could belong to the hidden or 
messenger sector of SUSY models \cite{candidates}. 
The most promising production mechanism proposed so far is the
enhancement of vacuum fluctuations of the $X$ field
during inflation by gravitational interactions \cite{ckr/kt}.
For $10^{12}~{\rm GeV}\lsim m_X \lsim 10^{13}$~GeV, this mechanism results
in a $X$-particle abundance close to the critical one.

\end{enumerate}

A common signature \cite{bbv} of all top-down models is the high 
photon/proton ratio $\gamma/p$. 
Since in the fragmentation process much more mesons
than baryons are produced, the ratio $\gamma/p$ should be $\gsim 1$ at
the highest energies. (The exact value depends on the fragmentation
model and on the
poorly known absorption length of UHE photons.)
If the UHE primaries originate from the decay of some superheavy CDM
particles, an additional signature is their
anisotropy \cite{bbv,dt98} reflecting the non-homogeneous distribution of the
CDM.

\section{Fragmentation spectrum of hadrons in SUSY-QCD}

The spectra of hadrons produced in deep-inelastic scattering and 
$e^+e^-$ annihilation are formed due to QCD cascading of the partons. 
In the Leading Logarithmic Approximation (LLA) which takes into
account $\ln(Q^2)$ terms this cascade is described by the 
Gribov-Lipatov-Altarelli-Parisi-Dokshitzer (GLAPD) equation~\cite{GLAPD}.
This approximation is not valid for Bjorken $x\ll 1$, when colour coherence
effects become important. A better approximation is the Modified LLA
(MLLA), which takes into account both $\ln(Q^2)$ and $\ln(x)$ terms as
well as angular ordering. In this section, we consider the
{\em limiting spectrum\/}~\cite{dkmt} that has been obtained as an
approximate solution to the MLLA evolution equations for small 
$x$. In fact, it describes well also the experimental data at $x\sim 1$. 

At large 
energies $\sqrt{s}\gsim 1$~TeV the production of supersymmetric 
particles might substantially change the QCD spectra. Apart from future 
experiments at LHC, supersymmetry (SUSY) might strongly reveal itself 
in the decays of superheavy particles: 
As long as the virtuality $Q^2$ remains much larger than the SUSY scale 
$M_{\rm SUSY}^2$, in the particle cascade initiated by the decay of a
$X$-particle not only usual particles but also their supersymmetric
partners participate. As we will see, the
fragmentation spectrum of hadrons changes considerably going from QCD
to SUSY-QCD~\cite{bk98b}. 
To obtain reliable results for UHECR fluxes, it is therefore 
necessary to calculate the spectra within SUSY-QCD.

We consider first the SUSY-QCD cascade in LLA. 
Neglecting terms proportional to $\alpha(Q^2)$ and 
keeping terms with $\alpha_s(Q^2) \ln(Q^2)$, 
the GLAPD equation can be written as~\cite{dkmt}
\be   \label{GLAP}
 \frac{\partial}{\partial\xi} D_A^B(x,\xi)=
 \sum_C\int_0^1 \frac{dz}{z} \: \Phi_A^C(z) D_C^B(x/z,\xi) -
 \sum_C\int_0^1 dz \: z \Phi_A^C(z) D_A^B(x,\xi) \,,
\ee
where $\Phi_A^B(z)$ is the splitting function \cite{split} characterizing the
decay $A\to B+C$.
Here $D_A^B$ is the distribution of partons $B$ inside the parton $A$
dressed by QCD interactions with coupling constant $\alpha(k_\perp^2)$, 
where $k_\perp$ is the transverse momentum and 
$x=k_{\parallel}/k_{\parallel}^{\rm max}$  is the longitudinal
momentum fraction of the parton $B$. The variable $\xi$
characterizes the maximum value of $k_\perp^2$ available in the
considered process ($k_\perp^2 < Q^2$),
\be
 \xi(Q^2)=\int_{\Lambda^2}^{Q^2} \frac{dk_\perp^2}{k_\perp^2} \:
                                 \frac{\alpha_s(k_\perp^2)}{4\pi} \:,
\ee
with $\Lambda\sim 0.25$~GeV as phenomenological parameter.

The supersymmetrization of Eq.~(\ref{GLAP}) is simple: each 
parton $A$ should be substituted by the supermultiplet which contains
$A$ and its superpartner $\tilde A$. We are mainly interested 
in the spectrum at small $x$, where gluons strongly dominate
the other partons. Therefore, it is a reasonable approximation to
consider only two partons, namely gluons $g$ and gluinos
$\lambda$, in the tree diagrams. 

After performing a Mellin transformation,
\be
 D_A^B(j,\xi) = \int_0^1 dx \: x^{j-1}  D_A^B(x,\xi) \,,
\ee
Eq.~(\ref{GLAP}) can be rewritten in matrix form, 
\be    \label{S}
 \frac{\partial}{\partial\xi} D(j,\xi)= H(j)D(j,\xi) \,,
\ee
where we have chosen as basis $(g,\lambda)$ and
\be   
 H(j) = \left( \begin{array}{cc}
               \nu_g(j) & \Phi_g^\lambda(j) \\
               \Phi^g_\lambda(j) & \nu_\lambda(j)
               \end{array}
        \right)
\ee
\be
 \nu_g(j) = \int_0^1 dz \left[ \left( z^{j-1}-z \right) \Phi_g^g(z)  
                               - \Phi_g^\lambda(z) \right]
\ee
\be      
 \nu_\lambda(j) = \int_0^1 dz \left( z^{j-1}-1 \right)
 \Phi_\lambda^\lambda(z) \:.
\ee
After diagonalization of $H(j)$, the eigenvalues of $H$ are
\be
 \nu_\pm = \frac{1}{2} \left( \nu_g + \nu_\lambda \pm
           \left[ \left( \nu_g - \nu_\lambda \right)^2
                  + 4 \Phi_g^\lambda\Phi^g_\lambda \right]^{1/2}
\right) \:.
\ee
In the limit $\omega=j-1\to 0$, the leading term $\nu_+$ is  given by
\be
 \nu_+ = \frac{4N_c}{\omega} -a + O(\omega)
\ee
with $a=\frac{11}{3} N_c=11.$ ($N_c=3$ is the number of colours.)

Dokshitzer and Troyan \cite{dt84} 
were able to express the limiting spectrum $D_{\rm lim}$ which
is a MLLA result as a function of the parameter
$a$ calculated in LLA and $b$, the constant of evolution of
$\alpha_s(k_\perp^2)$ in one-loop approximation,
$k_\perp^2 d\alpha_s(k_\perp^2)/dk_\perp^2=-b\alpha_s^2/(4\pi)$. 
Properly normalized, the function $D_{\rm lim} (l,Y)=xD_{\rm lim}(x,Y)$ gives
$\sigma^{-1}d\sigma/dl$ in the case of $e^+e^-$ annihilation and the
decay spectrum of $X$ particles, where $l=\ln(1/x)$,
$Y=\ln[\sqrt{s}/(2\Lambda)]$ and $\sqrt{s}$ is the c.m. energy of an $e^+
e^-$ pair or the mass $m_X$ of the superheavy decaying particle.
It is given by
\be
 D_{\rm lim}(l,Y) = K_{\rm lim} \,\frac{4C_F}{b}\,\Gamma(B)
      \int_{-\pi/2}^{\pi/2} \frac{d\tau}{\pi}\: e^{-B\alpha} 
      \left( 
        \frac{b}{8N_c}\:\frac{\sinh\alpha}{\alpha}\:\frac{Y}{y} 
      \right)^B  I_B (y)  
\ee
with $a=11$ and $b=b_{\rm SUSY}=9-n_f=3$ in SUSY-QCD. Furthermore,
$\alpha=\alpha_0 +{\rm i}\tau$, $\alpha_0={\rm arctanh}(2\zeta-1)$,
$\zeta=1-l/Y$ and $C_F=(N_c^2-1)/(2N_c)=4/3$. Finally,
$I_B$ is the modified Bessel function of order $B=a/b$,  and argument
\be
 y(\tau)=\left( \frac{16N_C}{b}\frac{\alpha}{\sinh\alpha}\:
                [\cosh\alpha+(1-2\zeta)\sinh\alpha] \: Y \right)^{1/2} \:.
\ee
In Fig.~2, the SUSY-QCD limiting spectrum is shown 
in comparison with the QCD limiting spectrum. The maxima of the SUSY
spectra are shifted to the right, and, since they are also narrower
than the QCD spectra, the SUSY maxima are dramatically higher. 

\vskip0.5cm
\unitlength1cm
\begin{picture}(15,6)
 \put (0.0,.7){\makebox(7,0)[b]{\epsfig{file={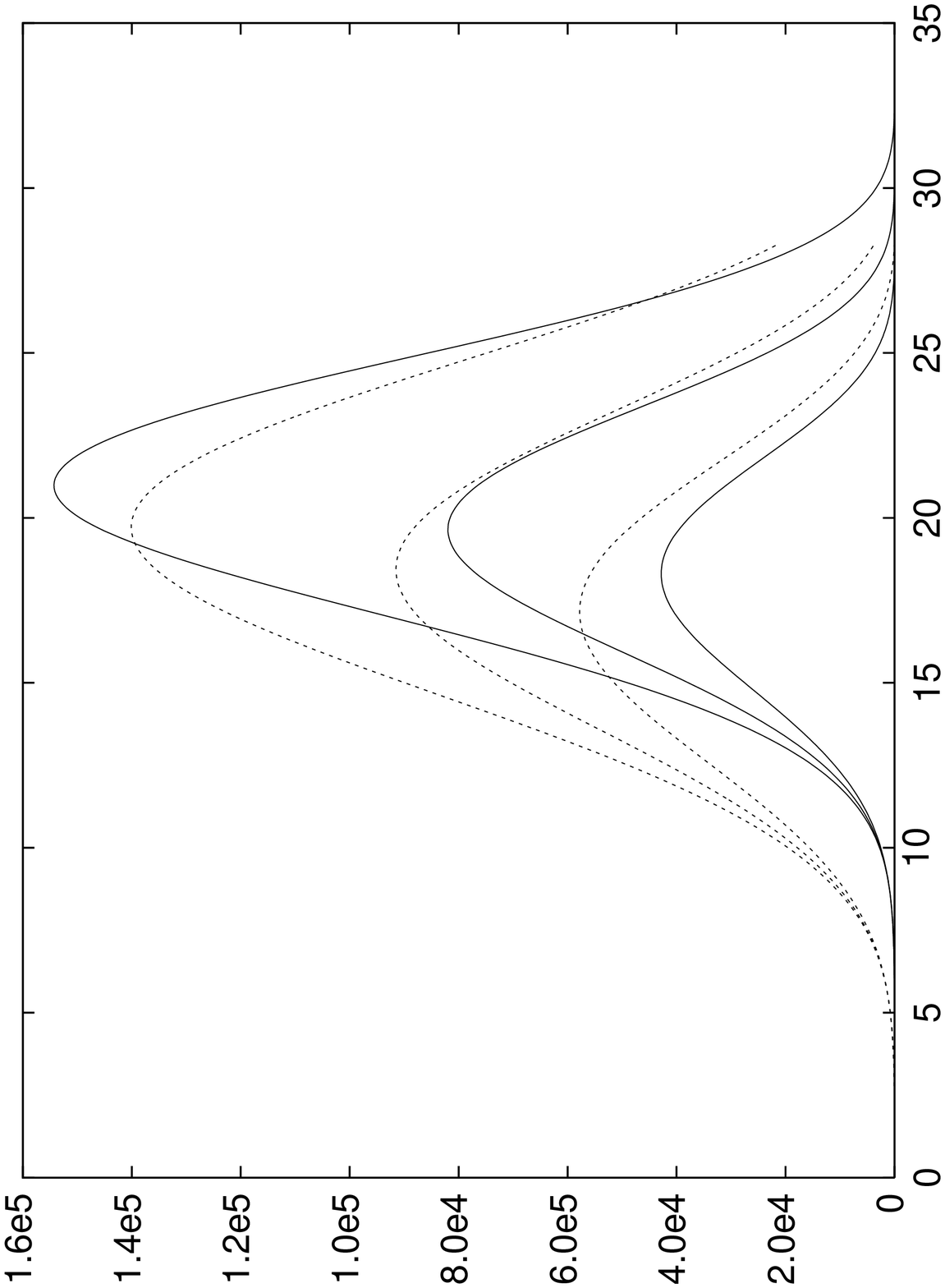},
                             height=7cm,width=5.0cm,angle=270}}}
 \put (3.2,0.2){\footnotesize $l=\ln(1/x)$}     
 \put (-0.4,2.8)
    {\begin{sideways} 
       {\footnotesize $D_{\rm lim}(l,Y)$}
     \end{sideways}}
 \put (8.3,.7){\makebox(7,0)[b]{\epsfig{file={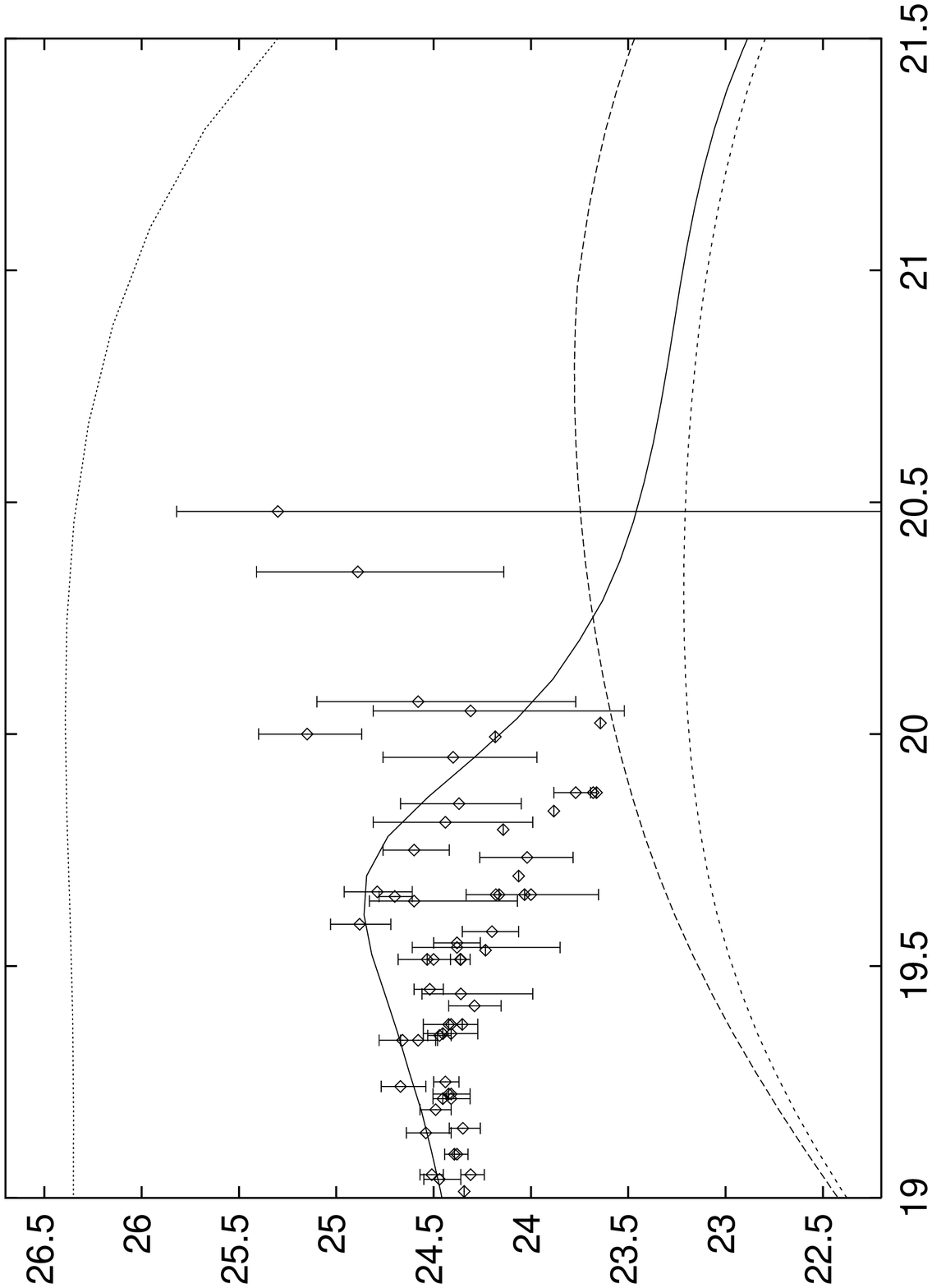},
                             height=7cm,width=5.0cm,angle=270}}}
 \put (11.5,0.2){\footnotesize $\log_{10}(E/{\rm eV})$}
 \put (8.0,1.1)
    {\begin{sideways} 
       {\footnotesize ${\rm log}_{10}(E^3 I)$,
                                 m$^{-2}$s$^{-1}$sr$^{-1}$eV$^2$}
     \end{sideways}}
 \put (14.,5.0){\footnotesize $I_\nu$}
 \put (14.,2.6){\footnotesize $I_\gamma$}
 \put (14.,2.05){\footnotesize $I_{p}$}
 \put (14.,1.5){\footnotesize $I_\gamma$}
\end{picture}

\def\baselinestretch{1.0}
{\footnotesize 
      \noindent
      Fig.~2 (left): Limiting spectrum $D_{\rm lim}(l,Y)$ for
      SUSY-QCD (solid lines) and QCD (dashed lines).
      The QCD spectrum is scaled up by a factor 30. 
      Both cases for $m_X=10^{12}$~GeV (bottom), $m_X=10^{13}$~GeV (middle)
      and $m_X=10^{14}$~GeV (top).
      
      \noindent
      Fig.~3 (right): Predicted fluxes from cosmic necklaces with 
      $r^2\mu=5\cdot 10^{27}$~GeV for $m_X=10^{14}$~GeV.
}

\def\baselinestretch{1.5}

\section{Application: UHECR fluxes from necklaces}
We present now the application of the SUSY-QCD limiting spectrum to the
calculation of UHECR spectra 
generated by the decay of superheavy particles~\cite{bk98b}.

Let us discuss first the problem of the normalization of spectrum.
The normalization constant $K_{\rm lim}$ cannot be calculated
theoretically and is normally found from comparison with experiment.
Since the spectrum changes dramatically going from QCD to SUSY-QCD, we
cannot use this value of $K_{\rm lim}$. Instead we use as  
normalization condition 
\be    \label{norm} 
 \int_0^1 dx \; xD_{i,{\rm lim}}(x,Y) = 2f_i \,,
\ee
where $i$ runs through $N$ (all nucleons) and $\pi^\pm$ and $\pi^0$
(charged and neutral pions), while $f_i$ is the fraction of energy
carried by the hadron $i$.
Note that the main contribution to the integral in Eq.~(\ref{norm})
comes from large values
$x\sim 1$, where the limiting  spectrum might have large uncertainties.
However, this is, in our opinion, the most physical way of normalization.
The numerical values of $f_i$ are unknown at large $s$. One can assume
that $f_\pi\approx 1-f_{\rm LSP}$, where $f_{\rm LSP}$ is the energy fraction
taken away by the lightest supersymmetric particle (LSP). According to
a simplified Monte-Carlo simulation~\cite{bk98}, $f_{\rm LSP}\sim
0.4$. For the ratio $f_N/f_\pi$ we use $\sim 0.05$ inspired by $Z^0$ decay.

Let us assume that the decay rate of $X$-particles $\dot n_X$ in the
extragalactic space does not depend on distance and time. Then taking
into account the energy losses of UHE protons and the absorption of
UHE photons due to pair production ($\gamma+\gamma\to e^+ +e^-$) on the
radio and microwave background, the diffuse flux of UHE protons and
antiprotons is
\be
 I_{p+\bar p} (E) = \frac{1}{2\pi} \: \frac{\dot n_X}{m_X}
                    \int_0^\infty dt_g D_N(x_g,Y)\frac{dE_g(E,t_g)}{dE} \:,
\ee
where 
$E_g(E,t_g)$ is the energy at generation time $t_g$ of a proton which has
at present the energy  $E$ and $x_g= 2E_g /m_X$.
Denoting the proton energy looses on microwave radiation by  $dE/dt=b(E,z)$, 
$dE_g/dE$ is given by 
\be  \label{looses}
  \frac{dE_g(E,z_g)}{dE} = (1+z_g) \exp\left[ \int_0^{z_g} \frac{dz}{H_0}
  \: (1+z)^{1/2} \left( \frac{\partial b(E,0)}{\partial E} \right)_{E=E_g(z)}
  \right] \:,
\ee
where $H_0$ is the Hubble constant and $z$ the redshift.
The diffuse spectrum of UHE photons can be calculated as
\be 
 I_\gamma(E) = \frac{\dot n_X}{\pi} \: \lambda_\gamma(E) \:
               \frac{1}{m_X} \int_{2E/m_X}^1 \frac{dx}{x} \: 
                D_{\pi^0} (x,Y) \:,
\ee
where $\lambda_\gamma$ is the absorption length~\cite{pb96} of a photon.
Finally, the diffuse neutrino flux is given by
\be    \label{nu}
 I_\nu(E) = \frac{3\dot n_X(t_0)}{\pi H_0 m_X} 
            \int_0^{z_{\rm max}} dz \: (1+z)^{\frac{3}{2}p}  
            \int_{\frac{4E(1+z)}{m_X}}^1 \frac{dx}{x} \: D_{\pi^\pm}
            (x,Y) \:,  
\ee
where $1+z_{\rm max}(E)\approx m_X/(4E)$. The neutrino flux depends
generally on the evolution of the sources,
\be  \label{n_X(t)}
 \dot n_X (t) = \dot n_X (t_0) \left(\frac{t_0}{t}\right)^{3+p} \:.
\ee

In Fig.~3, the spectra of UHE protons, photons, and neutrinos 
are shown together with experimental data for the model of cosmic
necklaces 
with $r^2\mu=5\cdot 10^{27}$~GeV$^2$ and $m_X=1\cdot 10^{14}$~GeV.
The proton flux is suppressed at the highest energies as compared with
the calculations of Ref.~[4], 
where the Gaussian SUSY-QCD spectrum was used.

\section{The LSP as UHE primary}
We have noted in Section 3 that usual particles as well as their
supersymmetric partners participate in a particle cascade as long as their 
virtuality $Q^2$  remains larger than the SUSY scale $M_{\rm SUSY}^2$.
However, when $Q^2$ reaches $M_{\rm SUSY}^2$, the supersymmetric
particles stop branching and decay to the LSP. Thereby, UHE LSP will
be produced in top-down models~\cite{bk98} (assuming $R$-parity is unbroken).
Since the LSP cannot be effectively accelerated in standard
astrophysical sources, the detection of UHE LSP in extensive air
shower (EAS) experiments would not only show that SUSY is realized by Nature
but would be also a clear signature for top-down models.

\subsection{Gluino as LSP}
Hadronic bound states of the gluino $\tilde g$ (which we call
generically $\tilde g$-hadrons $\tilde{G}$)  
were discussed already in the 80's as primary of the UHECR \cite{BI}.
Recently, this idea was revived by Farrar et al.~\cite{fa96} 
proposing as UHE primary the {\em gluebarino\/} $S=\tilde g uds$. 
A simple model that leads to the MSSM with gluinos as LSP was presented
by S. Raby~\cite{ra98}.

There are several arguments against a very light gluino with
${\mathcal{O}}(m_{\tilde g})=1-10$~GeV: First, the running of $\alpha_s$
gives~\cite{cf97} the bound $m_{\tilde g}\geq 6.3$~GeV. Second,
two searches for $\tilde g$-hadrons at Fermilab had negative
results~\cite{kTeV}. Third, $\tilde g$-hadrons are produced by CR in the 
earth atmosphere. In the case that the lightest gluebarino is the state 
$\tilde guud$ and stable, the light gluino
as LSP is excluded by the search for heavy hydrogen \cite{VO}.  
If the lightest gluebarino is neutral, as considered by Farrar et al., this
argument~\cite{VO} still work if $S$ forms a bound state with the nuclei.  
Thus, in our view, a very light gluino is disfavoured.
If the gluino is heavy enough (according Ref.~[20] 
$m_{\tilde g}
\gsim 50$~GeV), it will produce the standard missing energy signal in
accelerator experiments.
Therefore, the range $50~$GeV~$\lsim m_{\tilde g}\lsim 154$~GeV is
also excluded.

%
The interactions of UHE $\tilde{g}$-hadrons  were already
considered~\cite{BI} for the case of a glueballino.
Two values determine the interaction of a UHE $\tilde{g}$-hadron with
a nucleon. 
The first one is its total $\tilde{G}$N-cross-section which should be
proportional to its size. 
However, different estimates have been made in the literature: 
$\sigma \sim \alpha_s(m_{\tilde g})/ \mu^2 \sim 1$~mb~\cite{BI},
where $\mu$ is the reduced mass of $\tilde G$, 
$\sigma \sim \Lambda_{QCD}^{-2} \sim 10$~mb~\cite{mo98}, and 
$\sigma \sim (1-10)\, \sigma_{\pi N} \sim (3-30)$~mb~\cite{al98}.
The second important value is the average energy transfer $\langle y \rangle$ 
per scattering.
For the production of EAS in the atmosphere only interactions with
large $y$ are effective:
While a very light $\tilde{g}$-hadron would interact like a nucleon,
a heavy one behaves in the atmosphere like a penetrating particle and
should be distinguishable from a proton \cite{bk98,al98}.

\subsection{Neutralino as LSP}

In most SUSY models, the neutralino is the LSP and, in contrast to the
gravitino and gluino, it is also a viable DM candidate.   
Let us consider the interactions of the neutralino $\chi$ 
relevant for their detection~\cite{bk98}. Mainly two processes are important
for its interaction with matter, namely the neutralino-nucleon scattering
$\chi+N\to$ all and resonant production of selectron off electrons 
$\chi+e \to \tilde{e} \to$ all.  The first process is based on the
resonant subprocess $\chi+q \to \tilde{q} \to$ all and on neutralino-gluon
scattering. The latter subprocess is important, because for
high energies, and consequently for small scaling variable $x$, the gluon 
content of the nucleon increases fast.

The cross-sections of all subprocesses start to grow at 
energies $s\gg m_{\tilde q}^2$. The rise with $s$ is
caused by the decrease of $x_{\rm min}= m_{\tilde q}^2/s$ and
$x_{\rm min}= (m_{\tilde q} + m_q)^2/s$, and the corresponding increase of
the number of partons with sufficient momentum in the nucleon.
If squarks do not decay mainly into neutralino, 
{\it i.e.\/} $\Gamma_{\rm tot}\gg \Gamma(\tilde q_{L,R} \to q + \tilde\chi)$,
neutralino-gluon scattering gives the dominant
contribution to the total cross-section. 
At energies $s\ap 10^{10} \:$(GeV)$^2$ or $E_\chi \ap 5 \cdot 10^{18}$eV, 
the neutralino-nucleon cross-section is about
$10^{-35}-10^{-34}\:$cm$^2$ (for $m_{\tilde q}\sim 900$~GeV), 
{\it i.e.\/} slightly lower than the neutrino-nucleon cross-section. 

Let us consider now $\chi +e \to \tilde e\to$~all which is similar to the
Glashow 
resonant scattering $\bar\nu_e +e\to W \to \mu+\bar\nu_\mu$. 
The resonant energy of the neutralino is
\be
E_{\chi}=M_{\tilde e}^2/(2m_e)=9.8\cdot 10^8 (M_{\tilde e}/10^3~{\rm
GeV})^2~{\rm GeV}
\ee  
and the cross-section is given by the usual Breit-Wigner formula.
The resonant events are produced as a narrow 
peak at $E_{\chi}$ and give an unique signature for the neutralino.

\section{Conclusions}
Superheavy, metastable relic particle and cosmic necklaces are the two
most promising 
sources for top-down models. Their signature is the high photon/proton
ratio and the LSP as UHE primary. The UHECR flux produced by relic particles 
has additionally a small galactic anisotropy and, as most prominent signature,
no GZK-cutoff. For a reliable calculation of UHECR fluxes the knowledge of the
fragmentation spectrum of superheavy particles is necessary. We
presented the limiting spectrum in SUSY-QCD that differs drastically from
the QCD spectrum.

\section*{Acknowledgements}
This talk is based on work~\cite{bk98b,bk98} done together with
V.~Berezinsky whom I would like to thank for a very pleasant
collaboration. I am grateful to the Alexander von Humboldt-Stiftung for 
a Feodor-Lynen grant.


\end{document}